\documentclass[12pt]{article}
\usepackage{graphics}
\usepackage{amsthm}
\usepackage{amssymb}
\usepackage{amsmath}
\usepackage{amsfonts}
\usepackage{epic}
\usepackage{hyperref}
\usepackage{epsfig}
\usepackage{bm}
\usepackage{amscd}
\usepackage{amsmath,esint}

\theoremstyle{plain}

\theoremstyle{definition}

\def\be{\begin{equation}}
\def\ee{\end{equation}}
\smallskip

\begin{document}

\headsep=-0.5cm

\begin{titlepage}
\begin{flushright}
\end{flushright}
\begin{center}
\noindent{{\LARGE{Remarks on celestial amplitudes and Liouville theory}}}

\smallskip
\smallskip
\smallskip

\smallskip
\smallskip

\smallskip

\smallskip

\smallskip
\smallskip
\noindent{\large{Gaston Giribet}}
\end{center}

\smallskip

\smallskip
\smallskip

\centerline{Department of Physics, New York University}
\centerline{{\it 726 Broadway, New York, NY 10003, USA.}}

\smallskip
\smallskip

\smallskip
\smallskip

\smallskip

\smallskip

\begin{abstract}

\end{abstract}
The relation between celestial holography and Liouville field theory is investigated. It is shown that duality relations between different Selberg type integrals appearing in the Coulomb gas realization of Liouville correlation functions induce a series of relations between celestial amplitudes with shifted values of the operators dimensions $\Delta $.

This is a transcript of the talk delivered by the author at the Workshop on Celestial Holography and Asymptotic Symmetries, Santiago de Chile, March 4-6, 2024.
\end{titlepage}

\newpage


\section{Introduction}

Liouville theory \cite{Seiberg, Teschner, Nakayama} is a non-rational conformal field theory (CFT) connected to various problems in theoretical high energy physics, including the non-critical string, 2D string theory and black holes, 2D quantum gravity and random matrices, 3D Einstein gravity in AdS space, 4D $\mathcal{N}=2$ gauge theories, string theory on AdS$_3\times M$ backgrounds, among others. Recently, new applications of Liouville theory have been discovered; among the most salient ones we find its relation with the double-scaled SYK model \cite{Verlinde} and its relation with celestial amplitudes in gauge theory \cite{STZ, STZ2, Taylor}. In this talk, we will be concerned with the latter. 

In \cite{STZ}, the authors studied $n$-gluons maximally helicity violating (MHV) scattering amplitudes in Yang-Mills (YM) theory in the presence of a non-trivial dilaton background, and they constructed the corresponding celestial amplitudes. As usual, this is achieved by taking the Mellin transform with respect to the light-cone energy variables \cite{Pasterski3, Strominger, Pasterski}. What they found is that, when the celestial YM amplitudes are evaluated in the presence of a spherical dilaton shockwave, the corresponding CFT$_2$ observables can be expressed as $n$-point correlation functions of primary field operators in Liouville theory dressed with holomorphic currents. This seems to establish a novel relation between Liouville theory and 4D gauge theories, at least at classical level. 

In \cite{STZ2}, the relation between celestial amplitudes and Liouville correlators was studied beyond the classical level by exploring the 1-loop contributions. The conjecture was that, while tree-level approximation in YM theory corresponds to infinite central charge limit of Liouville theory, the subleading terms in perturbation theory could also admit a Liouville interpretation provided one conveniently relates the inverse of the Liouville central charge with the YM coupling constant $g_{\text{YM}}$ and the 1-loop $\beta$-function.

The connection between Liouville celestial amplitudes has also been explored in \cite{Taylor} in the context of supersymmetric YM theory coupled to the dilaton.

Here, we will analyze the Liouville theory realization of celestial amplitudes further. We will focus on the case of 4-point MHV gluon amplitudes. We will show that duality relations between different Selberg type integrals appearing in the Coulomb gas realization of Liouville correlation functions induce a series of relations between celestial amplitudes with shifted values of the operator dimensions $\Delta $. In section 2, we will review Liouville theory and collect notation. In section 3, we will revisit the computation of Liouville correlation functions in the free field approach. The connection to celestial amplitudes will be commented in section 4, where the relation between different amplitudes will be discussed. Section 5 contains some final remarks.

\section{Liouville theory}

Liouville theory \cite{Nakayama} is the conformal field theory (CFT) defined by the action
\begin{equation}
S[\phi ]=\frac{1}{4\pi }\int_{\Sigma } d^2 z \, \left(\partial\phi \bar\partial \phi + Q R \phi +4\pi \mu \, e^{2b\phi }\right)\label{action}
\end{equation}
where $\mu \in \mathbb{R}_{>0}$ is the so-called Liouville cosmological constant and $Q \in \mathbb{R}_{\geq 2}$ is the background charge. Conformal invariance requires the parameter $b\in \mathbb{R}$ and the background charge to satisfy the condition
\begin{equation}
Q=b+\frac 1b\, .
\end{equation}
This makes the exponential self-interaction term in (\ref{action}) to be a marginal operator with respect to the linear dilaton theory ($\mu =0$). $R$ is the scalar curvature of the 2-dimensional manifold $\Sigma $ on which the theory is defined. As we will be concerned with the conformal theory formulated on the sphere, $\Sigma=S^2$, we can locally chose holomorphic and anti-holomorphic coordinates ($z, \bar z$), with $\partial = \partial_{z}$, $\bar \partial = \partial_{\bar z}$, but taking into account the $\delta $-function contribution of the second term of (\ref{action}) at infinity when integrating over the entire space, e.g. when integrating over the zero mode of $\phi $ in the path integral calculation of correlation functions. In particular, this amounts to consider the Gauss-Bonnet theorem
\begin{equation}
\frac{Q}{4\pi }\int_{\Sigma } d^2 z\, R = Q\,\chi(\Sigma )\, ,\label{Q}
\end{equation}
with $\chi(\Sigma )$ being the Euler characteristic of $\Sigma $. Equivalently, the theory on the sphere ($\chi =2$) is defined by imposing the right asymptotic conditions in the limit $|z|\to \infty $; namely
\begin{equation}
\phi(z) \simeq -2Q\log |z|+\mathcal{O}(|z|^0)
\end{equation}
The topological invariant (\ref{Q}) is related to the fact that, at classical level, the theory results invariant under
\begin{equation}
\phi(z)\to \phi(z)+\phi_0 \, , \ \ \ \ \mu\to \mu \, e^{-2b\phi_0} \, . \label{cinco}
\end{equation}
That is to say, $\mu $ can be set to any arbitrary positive number by means of a shift of the zero mode $\phi_0$.

Liouville field $\phi $ is no exactly a scalar under the full conformal group. Under conformal map $z\to \omega (z)$ it transforms as follows 
\begin{equation}
\phi (z) \to \phi (z) -Q\log |\partial \omega | \, .
\end{equation}
The holomorphic and anti-holomorphic components of the stress tensor are 
\begin{eqnarray}
T(z) =-(\partial \phi )^2+Q\partial^2\phi \, , \ \ \ \ \ T(\bar z) =-(\bar\partial \phi )^2+Q\bar\partial^2\phi \, ,
\end{eqnarray}
whose operator product expansion (OPE) yields the central charge
\begin{equation}
c=1+Q^2\geq 25\, . \label{c}
\end{equation}

The primary states of the theory are created by exponential vertex operators \cite{Teschnervertex}
\begin{equation}
V_{\alpha} (z)= e^{2\alpha \phi(z)}\label{El7}
\end{equation}
with $\alpha $ being the so-called Liouville momentum. The corresponding states have conformal dimension
\begin{equation}
h_{\alpha}=\bar h_{\alpha} = \alpha (Q-\alpha ) \, .
\end{equation}
Normalizable states those with momentum
\begin{equation}
\alpha = \frac{Q}{2}+iP\ \ \ \text{with}  \ \ \ P\in \mathbb{R}\, ,
\end{equation}
which have dimension
\begin{equation}
h_{\alpha}=\bar h_{\alpha} = \frac{Q^2}{4}+P^2 \, .
\end{equation}
This implies that the theory has a gap in the continuous spectrum of normalizable states, namely 
\begin{equation}
h_{\alpha}=\bar h_{\alpha} \geq   \frac{c-1}{24}.
\end{equation}
The existence of the gap is linked to the way Liouville theory evades the obstruction imposed by the $c$-theorem: while its central charge (\ref{c}) is a continuous parameter, Liouville theory has no $SL(2,\mathbb{C})$-invariant normalizable vacuum state.

The states with momentum $\alpha $ and those with momentum $Q-\alpha $ have the same conformal dimension. Acting on normalizable states, this corresponds to the reflection $P\to -P$. At the fixed point $P =0$, two operators of conformal dimension $h_{\frac Q2}=\bar h _{\frac Q2} =\frac 14 Q^2$ exist; they are
\begin{equation}
V_{\frac Q2}(z) = e^{Q\phi (z)}   \ \ \ \text{and}  \ \ \ 
\hat{V}_{\frac Q2}(z) = \phi(z)\, e^{Q\phi (z)}
\end{equation}
Marginal operators ($h_{b^{\pm 1}}=\bar h _{b^{\pm 1}} = 1$) correspond to
\begin{equation}
V_{b}(z)= e^{2b\phi (z)}   \ \ \ \text{and} \ \ \ \ V_{\frac 1b}(z)= e^{{2}{b^{-1}}\phi (z)}\, .
\end{equation}
While the first of these two operators corresponds to the self-interaction term in (\ref{action}), the second can be thought of as a non-perturbative contribution which decouples in the classical limit $b\to 0$. To take the classical limit, it is convenient to define the rescaled quantities \cite{Higher}
\begin{equation}
\phi_c(z)=2b\, \phi(z) \, , \ \ \ \mu_c= 2\pi b^2 \mu
\end{equation}
and then send $b\to 0$. In that limit, the central charge (\ref{c}) tends to infinity ($c\sim 6/b^2$) and the stress tensor requires the renormalization $T_c(z)= b^2T(z)$. Here, we will be concerned with formulae for the theory at finite $b$, although the connection proposed with celestial amplitudes so far involved a semiclassical expansion, around $b\simeq 0$. 

The $n$-point correlation functions of primary operators on the Riemann sphere are defined as
\begin{equation}
\Big\langle \prod_{i=1}^nV_{\alpha_i}(z_i) \, \Big\rangle \, = \, \int_{\phi _{(\mathbb{CP}^1)}}\mathcal{D}\phi \, e^{-S[\phi ]}\prod_{i=1}^n e^{2\alpha_i\phi (z_i)}\, ,
\end{equation}
where we are omitting the normal ordering symbols and other decorations. In the next subsection we will review how these observables can be computed using the Coulomb gas approach, which is a crucial ingredient to make contact with the celestial amplitudes.

\section{Coulomb gas}

Liouville correlation functions can be computed in the path integral approach \cite{GL}. This amounts to first integrate over the zero mode $\phi_0 $, which yields the following integral expression
\begin{eqnarray}
\Big\langle \prod_{i=1}^nV_{\alpha_i}(z_i) \, \Big\rangle = \mu^s \Gamma(-s) \prod_{1\leq i<i'}^{n}|z_i-z_{i'}|^{-4\alpha_i\alpha_{i'}} \int_{\mathbb{C}^s} \prod_{r=1}^sd^2w_r \prod_{j=1}^{n}\prod_{t=1}^{s}|w_t-z_j|^{-4b\alpha_j} \prod_{1\leq l<l'}^{s}|w_l-w_{l'}|^{-4b^2} \nonumber
\end{eqnarray}
with the number of integrals being determined by
\begin{equation}
\sum_{i=1}^n \alpha_i = Q-s\, b\, ,\label{hhh}
\end{equation}
which uses that $\chi (\mathbb{CP}^1)=2$. In addition, one may resort to $SL(2,\mathbb{C})/\mathbb{Z}_2$ projective invariance and set $z_1=0$, $z_2=1$, $z_3=\infty $. The integration over $\phi _0$ produces the insertion of $s$ additional operators $\int d^2w_rV_{b}(w_r)$ as well as the divergent overall factor $\Gamma (-s)$. The $s$ operators are screening charges and can be physically understood as the contributions of the quanta that constitute the exponential potential in (\ref{action}). The divergent overall factor can be analyzed by expanding 
\begin{equation}
\Gamma (\varepsilon -s) \, \simeq \, \frac{(-1)^s}{\varepsilon\, s!} +\mathcal{O}(\varepsilon^0) \, , \ \ \ \ s\in\mathbb{Z}_{\geq 0},
\end{equation}
and combined with other $\Gamma $-functions coming from solving the integrals.

The integral expression above is obviously well defined for $s\in \mathbb{Z}_{\geq 0}$; analytic continuation is needed for the most general kinematic configurations, according to (\ref{hhh}). The integrals are defined over $\mathbb{C}$. The measure of each of them is $ d^2w_r = \frac i2\, dw_r d\bar{w}_r$, which can be separated in real and imaginary parts, i.e. $w_r = x_r + i y_r$, $\bar{w}_r = {x}_r - i {y}_r$. In order to integrate, first it is convenient to Wick rotate $x_r \to ix_r$ and then introduce a deformation parameter $\varepsilon $ by defining $| w_a |^2= -x_r ^2+y_r ^2+i\varepsilon $, which permits to avoid the poles at $x_r =\pm y_r $. Finally, one can define coordinates $x^{\pm }_r = \pm  x_r + y_r $ and integrate over $x^-_r$ while keeping $x^+_r$ fixed. Iterating this procedure and taking care of the combinatorics, one arrives to a closed expression for the particular case $n=3$. To analytically continue such expression, one first assumes $s\in \mathbb{Z}_{\geq 0}$ and, after having solved the integrals, which typically involves $\Gamma $-functions and its extensions, one continues to $s\in \mathbb{C}$. In the case $n=3$, the integral for any $s\in \mathbb{Z}_{\geq 0}$ reads \cite{DF}
\begin{eqnarray}
 \int_{\mathbb{C}^s} \prod_{r=1}^sd^2w_r \, \prod_{t=1}^{s} \left( \,|w_t|^{-4b\alpha_1} |w_t-1|^{-4b\alpha_2} \right) \prod_{1\leq l<l'}^{s}|w_l-w_{l'}|^{-4b^2} = \Gamma (s+1) \, \pi^s \left(\frac{\Gamma (1+b^2)}{\Gamma (-b^2)}\right)^{s}\nonumber \\
 \prod_{r=1}^{s}\frac{\Gamma (-rb^2)\, \Gamma (1-2b\alpha_1+(1-r)b^2)\, \Gamma (1-2b\alpha_2+(1-r)b^2)\, \Gamma (b\alpha_{1}+b\alpha_2-b\alpha_3+(r-1)b^2)}{
\Gamma (1+rb^2)\, \Gamma (2b\alpha_1+(r-1)b^2)\, \Gamma (2b\alpha_2+(r-1)b^2)\, \Gamma (-b\alpha_{1}-b\alpha_2+b\alpha_3+(1-r)b^2) }\nonumber
\end{eqnarray}
with $s=1+b^{-1}(1-\alpha_1-\alpha_2-\alpha_3)$. In the particular case $s=1$ it reduces to the Shapiro-Virasoro integral
\begin{equation}
\int_{\mathbb{C}}d^2w\, |w|^{-4b\alpha_1}|1-w|^{-4b\alpha_1}= \pi \prod_{i=1}^3\frac{\Gamma (1-2b\alpha_i)}{\Gamma (2b\alpha_i)}\, , \ \ \ \sum_{i=1}^{3}\alpha_i = 0\, .
\end{equation}
In the case $n=3$, $s\in \mathbb{Z}_{\geq 0}$, one can use that 
\begin{equation}
\Gamma (-s-\varepsilon ) = (-1)^{s}\, \frac{\Gamma (\varepsilon )\Gamma (1-\varepsilon )}{\Gamma (s+1-\varepsilon )}\, 
\end{equation}
and write the product of $\Gamma$-functions in terms of the well known $\Upsilon_b$ function, namely
\begin{equation}
\prod_{r=1}^{x}\frac{\Gamma (rb^2)}{\Gamma (1-rb^2)} \, =\frac{\Upsilon_b (bx+b)}{\Upsilon_b (b)} \, b^{x(b^2(x+1)-1)}\, ,
\end{equation}
and show that, after a proper analytic continuation in $x$, the right hand side of the integral formula above exactly matches the Dorn-Otto-Zamolodchikov-Zamolodchikov (DOZZ) formula \cite{DO, ZZ}. This is likely the most direct way to arrive to that formula. 

A very special set of correlators are the often-called resonant correlators. These are those involving states with the momenta obeying the condition
\begin{equation}
\sum_{i=1}^n \alpha_i = Q-m\, b-\tilde m\, b^{-1} \ \ \ \ \text{with} \ \ \ \ m, \tilde m\in \mathbb{Z}_{\geq 0}\, .
\end{equation}
These correlators exhibit poles and their residue are given by 
\begin{eqnarray}
&& \ \ \ \frac{\mu^m}{m!}\frac{\tilde\mu^{\tilde m}}{\tilde m!}\, \prod_{1\leq i<i'}^{n}|z_i-z_{i'}|^{-4\alpha_i\alpha_{i'}} \int_{\mathbb{C}^m} \prod_{r=1}^md^2w_r\int_{\mathbb{C}^{\tilde m}}\prod_{r=1}^{\tilde m}d^2v_l \, \prod_{r=1}^{m} \prod_{l=1}^{\tilde m}|v_l-w_{r}|^{-{4}} \nonumber \\ && \ \prod_{j=1}^{n} \prod_{t=1}^{m}|w_t-z_j|^{-4b\alpha_j} \, \prod_{j=1}^{n}\prod_{k=1}^{\tilde m}|v_k-z_j|^{-4\frac{\alpha_j}{b}}  \prod_{1\leq r<r'}^{m}|w_r-w_{r'}|^{-4b^2} \prod_{1\leq l<l'}^{\tilde m}|v_l-v_{l'}|^{-\frac{4}{b^2}}\label{resuena}
\end{eqnarray}
where
\begin{equation}
\tilde \mu = \frac{\Gamma (1-b^{-2})}{\pi\, \Gamma (b^{-2})} \left( 
\frac{\pi  \mu \, \Gamma (b^2)}{\Gamma (1-b^2)}
\right)^{b^{-2}}\,.
\end{equation}
This yields the correct Knizhnik-Polyakov-Zamolodchikov scaling of the Liouville correlators,
\begin{equation}
\Big\langle \prod_{i=1}^nV_{\alpha_i}(z_i) \, \Big\rangle \, \sim \, \mu ^{\delta }  \ \ \ \ \text{with}\ \ \ \  \delta ={1+b^{-2}-b^{-1}\sum_{i=1}^n\alpha_i}\,.
\end{equation}
which follows from (\ref{cinco}). Integral realization (\ref{resuena}) can be understood in terms of the insertion of the $m+\tilde{m}$ screening operators which are needed to screen the background charge $Q$ coming from the linear dilaton term in (\ref{action}). In this picture, the integrand in (\ref{resuena}) comes from the free field realization
\begin{equation}
\Big\langle e^{2\alpha_i \phi (z_i)} e^{2\alpha_j \phi (z_i)}\Big\rangle_{\text{free}} = |z_i - z_j |^{-4\alpha_i \alpha_j }\, ;
\end{equation}
this is analog to the Coulomb gas representation in the minimal models \cite{DF, DF2}. Alternatively, (\ref{resuena}) can be directly obtained from the path integral representation \cite{GL} starting with the Liouville action (\ref{action}) supplemented with the non-perturbative operator
\begin{equation}
\delta S[\phi ]=\tilde \mu \int d^2 z \,   e^{{2}{b}^{-1}\phi (z)}\, .\label{elbravo}
\end{equation}
Correlators involving the insertion of the operator (\ref{elbravo}) are precisely those that are related to celestial amplitudes \cite{STZ, STZ2}.

\section{Celestial amplitudes}

The relation between Liouville correlation functions and celestial amplitudes involves CFT$_2$ operators of the form \cite{STZ, STZ2}
\begin{equation}
O^{\pm ,\, a}_{\lambda }(z) = F_{\pm }(\Delta , \mu ,b) \, J^a_{\pm }(z)\, e^{2\sigma b\phi(z)},\label{O}
\end{equation}
which consists of a Liouville operator (\ref{El7}) of momentum $\alpha = b\sigma$ in direct product with a holomorphic current: $J^a_{-}(z)$ is a current of chiral weight $(-1,0)$, which is in the adjoint representation of the gauge group, while $J^a_{+}(z)$ is an affine Kac-Moody current of weight $(+1,0)$. The label $\lambda $ is a function of $\sigma $ and of the dimension of the celestial operators, $\Delta $; see (\ref{alpfa})-(\ref{Deltas}) below. The index $\pm $ refers to the helicity of the corresponding state in the celestial amplitude: $-$ corresponds to negative helicity gluons and $+$ corresponds to positive helicity gluons. $F_{\pm }(\Delta , \mu ,b)$ is a normalization factor,
\begin{equation}
F_{\pm }(\Delta , \mu ,b) = \left(  \frac{ \pi\mu \, \Gamma(b^2)}{\Gamma(1-b^2)}\right)^{\sigma}b^{-2b^{2}\sigma }\,\Gamma (2\sigma)\label{FFF}
\end{equation}
which, through $\sigma $, depends on $\Delta$ and so it contributes to the Mellin transform when going to the celestial basis.

The association of celestial amplitudes to CFT$_2$ correlators \cite{STZ} involving operators (\ref{O}) is motivated by the fact that celestial amplitudes yield an integral representation similar to the resonant Liouville correlators (\ref{resuena}) with $m=0$, $\tilde m = 1$, cf. \cite{Stieberger, Stieberger2, Stieberger3, Stieberger4}. The precise relation between CFT$_2$ correlators and gauge theory amplitudes proposed in \cite{STZ, STZ2} takes the following form
\begin{equation}
 \Big\langle O^{- ,\, a_1}_{\lambda_1 }(z_1)
O^{- ,\, a_2}_{\lambda_2 }(z_2)
\prod_{i=1}^n O^{+ ,\, a_i}_{\lambda_i }(z_i)
\Big\rangle \, \delta^{(n)}\left(\sum_{i=1}^n\lambda_i\right) =  \mathcal{M}_n (z_1,z_2,z_3,...z_n|\lambda_1-i,\lambda_2-i,\lambda_3,...\lambda_n)\label{La26}
\end{equation}
This formula connects $n$-point correlation functions in a CFT$_2$, including a Liouville factor, to MHV scattering amplitudes of $n$ gluons in celestial basis. As usual, going from the momentum space amplitudes to the celestial amplitudes amounts to write the 4-momentum of the $i^{\text{th}}$ gluon, $p^{\mu}_i$, in terms of the complex variable $z_i$; namely
\begin{equation}
p^{\mu}_i=(p^{+}_i,p^{-}_i,p^{2}_i,p^{3}_i)= \omega_i\, \left(1,|z|^2,\frac {z_i+\bar z_i}{2},\frac {z_i-\bar z_i}{2}\right) \, , \ \ \ \ i=1,2,...n\, ,
\end{equation}
where $\omega _i$ is the light-cone energy of the $i^{\text{th}}$ gluon, over which the Mellin tranform is to be taken. The variable conjugated to $\omega _i$ is the conformal dimension of the celestial operators, denoted by $\Delta _i$, which relates to $\sigma _i $ and $\lambda _i$ as described in (\ref{alpfa})-(\ref{Deltas}) below. 

On the left hand side of (\ref{La26}) we have decorated $n$-point Liouville correlation functions, and on the right hand side of it we have celestial MHV amplitudes with mostly plus signature and in presence of a non-trivial background: Unlike the standard celestial amplitudes, the one appearing in (\ref{La26}) are defined in the presence of a specific dilaton background; see \cite{STZ, STZ2} for the details. This is related to the fact that the first two arguments of $\mathcal{M}_n$ are evaluated on the shifted values $\lambda_{1,2}-i$. To see how this is related to the non-trivial background field, let us take a closer look at the dictionary between the Liouville and the celestial variables: As observed, operators (\ref{O}) involve a Liouville vertex operator corresponding to a light state
\begin{equation}
\alpha_i = b\sigma_i \, , \ \ \ \ \ i=1,2,3,...n \label{alpfa}
\end{equation}
i.e. these are states whose momenta tend to zero in the semiclassical limit $b\to 0$. A non-perturbative (finite-$b$) proposal for the relation between the Liouville momentum $\alpha_i$ and the celestial operator $\Delta _i$ is
\begin{equation}
\alpha_i=\frac{Q}{2} \left( 1-\sqrt{1-2Q^{-2}(\Delta_i -\epsilon_i )}\right)
\end{equation}
with $\epsilon_i = \pm 1$, depending on the helicity of the state. In the semiclassical limit, this yields 
\begin{equation}
\sigma_i \simeq \frac 12 (\Delta -\epsilon_i )+\mathcal{O}(b)\, . 
\end{equation}
In \cite{STZ, STZ2} the authors consider mostly plus $n$-point MHV amplitudes with $\epsilon_{1,2}=-1$, $\epsilon_{i\geq 3}=+1$. The relation between the Liouville momenta $\alpha_i=b\sigma_i$ and the parameters $\lambda_i$ is
\begin{eqnarray}
\sigma_1=\frac{1+i\lambda_1}{2}\, , \ \ \ \ \
\sigma_2=\frac{1+i\lambda_2}{2}\, , \ \ \ \ \
\sigma_{i\geq 3}=\frac{i\lambda_{i\geq 3}}{2}\, ,\label{lambdas}
\end{eqnarray}
This satisfies the constraint
\begin{equation}
\sum_{i=1}^n\sigma_i = 1 \, ,
\end{equation}
which is nothing but the charge saturation condition (\ref{hhh}), together with the celestial amplitude conservation condition  
\begin{equation}
\sum_{i=1}^n\lambda_i = 0\, .
\end{equation}
This corresponds to the following values of the conformal dimension of the operators in the celestial amplitudes
\begin{equation}
\Delta_1= i\lambda_1 \, , \ \ \ \ \ \Delta_2=  i\lambda_2 \, , \ \ \ \ \ \Delta_{i\geq 3}=  1+i\lambda_{i\geq 3}\, .\label{Deltas}
\end{equation}
That is to say, two of the states involved in the $n$-point amplitude have dimensions $\Delta$ shifted by $-1$ with respect to the principal series value $\Delta \in 1+i\mathbb{R}$ commonly considered in celestial holography, cf. \cite{Stieberger, Stieberger2, Stieberger3, Stieberger4}. 

The main point we want to make here is that the relation between CFT$_2$ correlators and celestial amplitudes, if correct, would ipso facto imply a series of relations among celestial amplitudes. The latter would follow from duality relations that the Selberg type integrals obey. In celestial holography, integrals similar to those of resonant $n$-point correlator (\ref{resuena}) with $m=0$, $\tilde m =1$ are known to appear. Considering the values (\ref{lambdas}) this yields the integral
\begin{eqnarray}
{\tilde\mu}\, \,\Omega _b(z_1, z_2, ...z_n|\lambda_1-i, \lambda_2-i, \lambda_3, ...\lambda_n)\, \int_{\mathbb{C}} d^2v\, |v|^{-2(1+i\lambda_1)}|1-v|^{-2(1+i\lambda_2)} \prod_{j=3}^{n} |v-z_j|^{-2i\lambda_j}  \label{rutera}
\end{eqnarray}
multiplied by a factor
\begin{eqnarray}
\Omega _b(z_1, z_2, ...z_n|\lambda_1-i, \lambda_2-i, \lambda_3, ...\lambda_n) = \prod_{j=1}^{n} |z_{j}|^{-ib^2(1+i\lambda_1)\lambda_{j}}|1-z_{j}|^{-ib^2(1+i\lambda_2)\lambda_{j}}
 \prod_{3\leq i<i'}^{n}|z_i-z_{i'}|^{b^2\lambda_i\lambda_{i'}} \nonumber
\end{eqnarray}
which in the semiclassical limit disappears, namely $\Omega _0(z_1, z_2, ...z_n|\lambda_1, \lambda_2, ...\lambda_n)=1$. Interestingly, these integrals obey abstruse duality relations. Specifically, this implies that the celestial amplitudes studied in \cite{STZ, STZ2}, which have momenta (\ref{Deltas}), will necessarily come along with other amplitudes involving other values of $\Delta_i$, including states with $\mathbb{R}\text{e}(\Delta )\leq 0$ and $\mathbb{R}\text{e}(\Delta )\in \mathbb{Z}_{> 1}$.

For example, we have the duality formula \cite{BF, FL, FL2}
\begin{eqnarray} 
&&\int_{\mathbb{C}^{m}}\prod_{i=1}^{m}d^2w_i \prod_{1\leq i <i'}^{m} |w_i - w_{i'}|^2 \prod_{i=1}^{m}\prod_{j=1}^{m+q+1} |w_i - z_j|^{2\eta _{j}} = \pi^{m-q}\, \frac{\Gamma(m+1)}{\Gamma(q+1)}\frac{\Gamma(-m-\sum_{j=1}^{m+q+1}\eta_j)}{\Gamma(1+m+\sum_{j=1}^{m+q+1}\eta_j)}   \nonumber \\
&&\, \prod_{j=1}^{m+q+1} \frac{\Gamma(1+\eta_j)}{\Gamma(-\eta_j)} \prod_{1\leq j <j'}^{m+q+1}|z_j - z_{j'}|^{2+2\eta_j+2\eta_{j'}} \int_{\mathbb{C}^q}\prod_{l=1}^{q}d^2w_l \prod_{1\leq l<l'}^{q} |w_{l}-w_{l'}|^2  \prod_{l=1}^{q} \prod_{j=1}^{m+q+1} |w_l - z_j|^{-2-2\eta_j} . \nonumber \\ \label{La}
\end{eqnarray}
For our purpose, it is sufficient to focus on the 4-point function ($n=4$). In the formula above, consider the case $m=1$, $m+q+1=n-1$ with $n=4$ and $z_4=\infty $, and so $q=1$ as well. This yields
\begin{eqnarray}
\eta_1&=&-1-i\lambda_1 \, , \ \ \ \ \ \eta_2=-1-i\lambda_2 \, , \nonumber\\
\eta_3&=&-i\lambda_3 \, ,  \ \ \ \ \ \ \ \ \ \ \eta_4=-i\lambda_4 \, .\nonumber
\end{eqnarray}
Since $m=q=1$, (\ref{La}) turns out to be an identity between two 4-point integrals of the form (\ref{rutera}). In other words, a Liouville resonant correlator with a single screening $\int d^2vV_{b^{-1}}(v)$ results in an other correlator of the same type but having changed
\begin{equation}
\alpha_i \to \frac{b}{2}-\alpha_i \, , \ \ \ \ \ i=1,2,3,4.\label{transformer}
\end{equation}
Notice that transformation (\ref{transformer}) is not the Liouville reflection that leaves $h_{\alpha }$ and $\bar h_{\alpha }$ invariant, but it rather corresponds to the change $h_{\alpha}\to h_{\alpha}+\frac 14$. One could have naively expected the relation (\ref{La}), which corresponds to the transformation (\ref{transformer}), to merely realize the shadow transform of the celestial operators. However, this is not what happens; transformation (\ref{La}) rather corresponds to a less obvious relation between states. While for states of positive helicity shadow transform would act by changing $\lambda_{3,4}\to -\lambda_{3,4}$, transformations (\ref{transformer}) produces the changes
\begin{eqnarray}
&i\lambda_1 \to -1-i\lambda_1 \, , \ \ \ \ \ i\lambda_2 \to -1-i\lambda_2 \, ,\\ 
&i\lambda_3 \to +1-i\lambda_3 \, , \ \ \ \ \ i\lambda_4 \to +1-i\lambda_4 \, .
\end{eqnarray}
In terms of the dimensions, this corresponds to 
\begin{equation}
\Delta_1\to -1-\Delta_1 \, , \ \ \ \Delta_2\to -1-\Delta_2 \, , \ \ \ \Delta_3\to 3-\Delta_3 \, , \ \ \ \Delta_4\to 3-\Delta_4 \, . 
\end{equation}

This means that the Coulomb gas representation of the amplitudes implies direct relations among celestial gluon amplitudes with different values of dimensions $\Delta_i$. Now, the meaning of these exotic shifts is unclear and this can either be seen as a consistency check for the CFT$_2$ realization or as a hint of new relations among observables.

\section{Remarks}

Before concluding, we would like to make some remarks about finite-$b$ duality relations between Coulomb gas integrals. It was suggested in \cite{STZ2} that the relation between celestial YM amplitudes and Liouville correlators persists at quantum level. In the Liouville theory side this corresponds to $\mathcal{O}(b^2)$ corrections, i.e. $1/c$ effects. The main idea in \cite{STZ2} is that, while tree-level approximation in YM theory corresponds to the $b\to 0 $ limit of Liouville theory, the subleading terms in perturbation theory also admit a Liouville interpretation provided one relates $b$ with the YM coupling constant $g_{\text{YM}}$ and the 1-loop $\beta$-function. Aiming at exploring this connection further, nothing prevents us from considering the finite-$b$ regime. In order to do so, let us consider a different duality relation between multiple Selberg type integrals. In contrast to the duality (\ref{La}), now we will be involved with a relation between integrals that is both non-diagonal and non-perturbative: it is non-diagonal in the sense that it mixes different momenta $\lambda _i$, and is non-perturbative in the sense that the relation between variables diverges in the limit $c\to \infty $.

Consider the 4-point Coulomb gas integral with $z_1=0$, $z_2=1$, $z_3=\infty $, and $z_4$ being an arbitrary point on $\mathbb{C}$. Then, the following identity holds
\begin{eqnarray}
&&\int_{\mathbb{C}^s} \prod_{r=1}^sd^2w_r \, \prod_{t=1}^{s} \left( \,|w_t|^{-4\sigma_1} |w_t-1|^{-4\sigma_2} |w_t-z_4|^{-4\sigma_4} \right) \prod_{1\leq l<l'}^{s}|w_l-w_{l'}|^{-{4}{b^{-2}}}
= \nonumber \\ &&\  
\left(-\frac{\Gamma(1+b^{-2})}{\Gamma(-b^{-2})}\right)^{s-m}
\prod_{r=1}^{s-m}\frac{\Gamma(2\sigma_4-rb^{-2})}{\Gamma(1-2\sigma_4-rb^{-2})} \prod_{l=0}^{s-m-1}\prod_{j=1}^{3} \frac{\Gamma (1-2\sigma_j-lb^{-2})}{\Gamma (2\sigma_j+lb^{-2})}
\nonumber \\ &&  \ 
\int_{\mathbb{C}^m} \prod_{r=1}^md^2w_r \, \prod_{t=1}^{m} \left( \,|w_t|^{-{4\tilde{\sigma}_1}} |w_t-1|^{-{4\tilde{\sigma}_2}} |w_t-z_4|^{-{4\tilde{\sigma}_4}} \right) \prod_{1\leq l<l'}^{m}|w_l-w_{l'}|^{-{4}{b^{-2}}}\label{nodiagonal}
\end{eqnarray}
with
\begin{equation}
s=b^2+1-b(\alpha_1+\alpha_2+\alpha_3+\alpha_4)\, ,
\end{equation}
together with
\begin{eqnarray}
\tilde{\sigma}_1 &=& \frac{1}{2b^2} + \frac{\sigma_1 - \sigma_2 - \sigma_3 + \sigma_4+1}{2} \, , \ \ \ \ \tilde{\sigma}_2 = \frac{1}{2b^2} - \frac{\sigma_1 - \sigma_2 + \sigma_3 - \sigma_4-1}{2} \, ,  \\
\tilde{\sigma}_3 &=& \frac{1}{2b^2} - \frac{\sigma_1 + \sigma_2 - \sigma_3 -\sigma_4-1}{2} \, , \ \ \ \ \tilde{\sigma}_4 = -\frac{1}{2b^2} + \frac{\sigma_1 + \sigma_2 + \sigma_3 + \sigma_4-1}{2} \, ,
\end{eqnarray}
and
\begin{equation}
m=1+b^2-b\sum_{i=1}^{4}\tilde{\alpha}_i=-2b\alpha_4\, .\label{cincuentona}
\end{equation}
This also yields 
\begin{equation}
\tilde{\sigma}_4=-\frac{s}{2b^2}\, ,\label{R51}
\end{equation}
which corresponds to a degenerate state for $s\in \mathbb{Z}_{>0}$.The degenerate representations are those of the form
\begin{equation}
\tilde \sigma_4= -\frac{r}{2} - \frac{s}{2b^2}\,  , \ \ \ \text{with} \ \ \ r,s\in\mathbb{Z}_{\geq 0}\, .
\end{equation}

Integral relation (\ref{nodiagonal})-(\ref{R51}) implies a non-trivial relation between 4-point functions of different sets of momenta. This has already been successfully applied to the Liouville description of gauge theories in \cite{GiribetAGT}, where the $SO(8)$ symmetry of the $SU(2)$ Seiberg-Witten theory with 4 flavors was studied in the context of AGT correspondence \cite{AGT}. Here, we would like to ask a similar question but in a new framework: what does the duality formula (\ref{nodiagonal})-(\ref{R51}) mean in the context of the Liouville realization of celestial amplitudes proposed in \cite{STZ, STZ2}?

Condition $\sum_{i=1}^4\lambda_i=0$ in the celestial amplitudes implies $r=0$, $s=1$. And defining $2\tilde\sigma_i=\epsilon_i+i\tilde \lambda_i$, with $\tilde\lambda_i\in\mathbb{R}$ for $i=1,2,3$, we have the following relations between variables of different correlators
\begin{eqnarray}
i\tilde\lambda_{1} = \frac{1}{b^2}-i(\lambda_{2}+\lambda_{3})\, , \ \ \ \ i\tilde\lambda_{2} = \frac{1}{b^2}-i(\lambda_{1}+\lambda_{3})\, , \ \ \ \ i\tilde\lambda_{3} = \frac{1}{b^2}-i(\lambda_{1}+\lambda_{2})\, . 
\end{eqnarray}
This relation is non-perturbative as it involves a $1/b^2$ term. In addition, this requires the analytic extension of the multiple integral formula above to complex values of $m$. Whether there is a way of understanding such a relation in the context of the investigation initiated in \cite{STZ} remains an open question.

\[\]

The author thanks Laura Donnay, Hern\'an Gonz\'alez, and Francisco Rojas for discussions. He also thanks the organizers and the participants of the Workshop on Celestial Holography and Asymptotic Symmetries, organized in Santiago de Chile, March 4-6, 2024.

\end{document}